\documentstyle[12pt]{article}
\begin{document}
\begin{titlepage}
\begin{minipage}{13cm}
\begin{flushright}
\begin{tabular}{l}
TPI--MINN--93/13-T\\
UMN-TH-1150/93
\end{tabular}
\end{flushright}
\vspace{1cm}
\begin{center}
{\LARGE QCD Sum Rules: The Second Decade}\\[5pt]
{Talk at the  Conference "QCD -- 20 Years Later"}\\
{June 9 -- 13, 1992, Aachen, Germany}\\
\vspace{2cm}

{\large M.A.\ Shifman}\\
\vspace{0.5cm}
{\large\em Theoretical Physics Institute, University of Minnesota,}\\
{\large\em Minneapolis, MN 55455, USA}\\[2pt]

\bigskip

{\large \today}\vspace{2cm}

\end{center}
{\small
\noindent   }

\vspace{2cm}

\end{minipage}
\end{titlepage}

\section{Introduction}

My task in this talk, as I understand it, is two-fold.  First, to describe
the basic idea of the QCD sum rule approach which was -- and still is
-- one of the most productive tools for calculating hadronic
parameters from Quantum Chromodynamics. Second, I would like to
try to convey the historic flavor of the exciting time when the theory
was making its first steps, with a euphoric hope (you
could feel it in the air)  of an imminent solution of the confinement
problem.

To make a proper perspective it is, perhaps, worth starting from the
second point. It seems fair to say that QCD was born
after the talk of Gell-Mann and Fritzsch \cite{1} (see also ref. \cite{2}
) in
which the color-octet gluons were introduced. The next step is
certainly the discovery of asymptotic freedom in 1973 \cite{3}. In
the
first few years, roughly speaking till 1975, the theorists' attention
was almost totally focussed on hard processes in which the
short-distance physics plays the dominant role. This topic became
hot and
fashionable,  piles of new papers appearing daily. The main
achievement of this period is  that people learned how to reliably
isolate the short distance contributions governed by the small
coupling constant and how to generalize  electrodynamical
perturbative calculations to non-Abelian theories.

A recent paper of Polyakov \cite{4} presenting his understanding of
the
development of  our field in the seventies is entitled "A View from
the Island" which gives a good idea of our place in the scientific
process in Moscow in those days.  The isolation was almost complete,
and we could not compete with  Western theorists in most
fashionable directions where the results seemed to be on the surface,
for obvious reasons.

Because of the total censorship preprints and journals from the West
used to come with enormous delays, and our own papers could be
submitted to the Western journals only after a complicated
procedure of getting clearance from half a dozen of different
instances,  typically a waste of  a few months.  Publication of
preprints was  an adventure due to bureaucratic limitations.  For
instance, one and the same paper could not be issued as a preprint
and, then, in a journal, and the preprint version could not be longer
than 25 typewritten pages (or 35, I do not remember exactly).  So we
had a lot of "fun" trying to muddle and deceive our censors  by
making different titles in  the preprint and journal versions of one
and the same work, changing the order of the authors or  serializing
preprint publications  like a detective story in a popular newspaper
(unlike the latter case, though, we had to make  an impression  that
each successive part is not connected to the previous).  Quite often
these tricks created a terrible mess, to say nothing about wasted
efforts.  Occasionally, in the most
important cases, one would risk to bypass the standard procedure by
using different "illegal" channels, mostly our Western friends. By the
way,  any contacts with the latter were also severely constrained.
There is a famous story about one of the scientific bosses in Dubna
who was instructing the Soviet participants of a conference before its
opening. He said: "Well..., we had to organize this conference, and
even invite some foreigners. Unfortunately, to my deep
regret, this time it will be impossible to completely avoid
contacts..."
Conferences in the West were open for a handful of  specially
selected,
through a humiliating procedure; and  even those few which took
place  in  the
Soviet Union were not always accessible. I remember,  for instance,
that one of my colleagues from ITEP and I were not granted
permission to participate in "Neutrino --77" in Baksan.

This is a long saga, and I could easily speak for an hour on this topic,
but now it is  clearly time to stop. To make the long story short I will
only say that making our results known was a difficult, nervous and
time-consuming part of our job. This largely predetermined the
choice of topics we could work on and formed in the ITEP theory
group and elsewhere a very special atmosphere, now gone forever.

In the seventies ITEP had one of the best groups in the world, an
excellent collection of enthusiasts whose attitude to physics was
totally "non-commercial".  People were always eager  to discuss with
each other every interesting scientific question  emerging  during the
seminar talks or  elsewhere, and these discussions quite often would
last till midnight.  You could easily find  experts in any conceivable
field or  direction who would  gladly share  their knowledge with
you. Our  common enemies  and common isolation created, as a
counterreaction, very strong  friendly and scientific ties, as the only
way of survival,  and helped develop protective values; the tacit
understanding that good physics was above all was among these
values.

The only drawback I can think of now, in retrospective, is the
general
negative attitude to field theories in the very end of sixties and
the beginning of seventies when I just appeared in ITEP. The reason
is obvious,
of course: the  influence of Landau and his discovery of the
zero-charge property in QED \cite{Land} -- the influence which was
alive and
very strong in the ITEP theory group in those days. The attitude to
the
field theory as to something absolutely not serious was so deeply
rooted
that the fact of the {\em anti}screening of the gauge coupling
constant in
non-abelian
theories which was reported in ITEP at least twice
\cite{Tere},
\cite{Khri} in the {\em sixties} has not been appreciated and
recognized
\cite{Vany}.
A restructuring in minds started only after the very same fact, the
surprising fall-off of the gauge charge at large distances, became
known
from ref. \cite{3}.

So, our start was relatively slow.
By 1974, however, we were fully submerged in this subject,
and shortly after  it became clear that the
cavalry attacks do not help to solve the
problem of confinement and that the wide-spread expectations of a
Messiah who would come soon and teach us the mystery of the
solution had to be tempered. The fact that instantons \cite{5}, a
beautiful
theoretical construction which was a breakthrough in the qualitative
understanding of the QCD vacuum \cite{6}, turned out to be useless
in the
quantitative sense because of the infrared divergences was a serious
blow. So we adopted a less ambitious approach (by "we" I mean
Valya Zakharov, Arkady Vainshtein and myself). The idea was to
start from short distances where the quark-gluon dynamics was
essentially perturbative and we felt ourselves on a firm ground, and
then to extrapolate to larger distances (where the hadronic states are
presumably formed) including  non-perturbative effects "step by
step" and  using some kind of an approximate procedure  to extract
information on  hadronic properties. Of course, this idea was rather
vague at first, the program as we know it now has been crystallizing
gradually.

It is rather difficult to identify the work which, for us,
was  the first crucial  step. With some reservations
I might say that the first step has been done in ref. \cite{7}.
Perhaps now the sum rule for the charmed
particle photoproduction obtained in \cite{7}  does not seem very
impressive but this analysis carried important elements which where
later laid  in the foundation of the sum rule method. A spectacular
success came after we united our efforts with V. Novikov, L. Okun
and M. Voloshin. It turned out that a whole variety of the
charmonium parameters are predictable essentially from pure
duality, and for about a year  we played the game of  constructing
the  charmonium widths and mass ratios from simple numbers. In
1977 a review report \cite{8} was submitted. At about that time it
became clear that the progress was limited; the method presented in
\cite{8} could not be reliably generalized to such typical
representatives
of the hadronic family as, say, $\rho$ mesons or nucleons. And the
desire
to get access to these hadrons was strong.

The remainder of this story, including its culmination -- introduction
of the gluon condensate \cite{9} in fall 1977 -- is described
elsewhere
\cite{10}.  The basic elements of the approach were already visible in
this
first work, although some new important findings, like e.g.
borelization, came  somewhat later. We worked at a feverish pace for
the whole academic year, accumulating a large number of results for
the hadronic parameters with the accuracy which we could never
expect
beforehand. In late spring 1978 the question of how all this wealth
could  be published became of prime concern to us. As usual, we had
a
lot of funny adventures (they can hardly be understood by the
Western physicists) in preparing the manuscript, typing it and
issuing preprints.  Needless to say, it was a serial publication, see
above.  As for the  journal article, Nuclear Physics was a natural
candidate, but previously we had bad luck with this journal: our
paper on penguins \cite{11} was buried in the editorial office for
more
than two years. We were too tired, however, to invent anything new
and decided to try our luck again. The report occupying the whole
issue of Nuclear Physics \cite{12} appeared a year later.

\section{The Main Idea}

The lagrangian of Quantum Chromodynamics is built from gluons and
quarks. At short distances these degrees of freedom fully describe
the dynamics, while at large distances where the hadrons are formed
the quark-gluon description becomes irrelevant: the Green functions
are strongly distorted  and nobody knows how to calculate them
from first principles. The hope which lies behind the approximate
method developed in [12] is as follows. Let us assume that there
exists a transitional domain of distances (stretching up to the sizes
of the classical states with small spins, like $\rho$)
where the effect  of the infrared component of the
quark and gluon Green functions can be parametrized in terms of a
few vacuum condensates.
If this assumption is correct then a systematic approach can be
developed allowing one to calculate the parameters of these classical
states. The technical basis of the approach is the Wilson operator
expansion (OPE) \cite{13} , a construction very close in spirit and
ideally
suiting our  purposes. Indeed, the essence of OPE is separation of all
field fluctuations in scales: the small scale fluctuations are accounted
for explicitly in the coefficient functions; the large scale fluctuations
are referred to the matrix elements of a set of (local) operators.
Generally speaking, this set is infinite.

If the ultimate theory of hadrons existed it should be able to answer
in full two questions: (i)  what are the values of the coefficient
functions, and (ii) what are the values of the matrix elements from
the infinite set mentioned above. Given the exact answers one can
readily extract from them the exact and complete information about
all hadronic parameters.

Will the exact answers be ever known? Time will show.  Our current
abilities are much more modest. The coefficient functions are,
obviously, approximately calculable as an expansion in the small
running coupling constant. In principle, they include both
perturbative and non-perturbative parts \cite{14}, but in practice we
have to limit ourselves to several leading terms in the perturbative
expansion.  As for the matrix elements, these quantities are
essentially non-perturbative, and can not be calculated in the
present-day QCD.  Therefore, the success of the method crucially
depends on whether or not a finite truncation of the infinite set of
operators is sufficient to ensure a reasonable accuracy in the domain
where the classical hadrons are formed. If the convergence of OPE is
good, and the first few terms in the expansion describe the
correlators under consideration up to distances of order of the $\rho$
meson size then the corresponding  few matrix elements can be just
parametrized. In this way we arrive at a workable substitute to the
ultimate theory, which, if does not make us completely happy, still
allows to investigate the important regularities of the hadronic
family in a model-independent way without postponing the task till
indefinite future.

The closest analogy I can think of  to this method is the propagation
of external objects injected at $\tau$=0 in a medium. The medium
structure is complicated and essentially we have no idea how to
obtain it at the microscopic level.  The motion of the injected objects
at $\tau\rightarrow\infty$ depends, generally speaking, on
unknown details of what is going on in the medium reacting on the
presence of the external objects.  If, however, the characteristic
frequency $\omega$ of the external objects  is much larger than that
of the medium,  we can consider the propagation of the objects
during the time interval $\tau$ = several units$\times\omega^{-1}$.
This is a large time interval with regards  to the external objects, so
that they have enough time to form a stationary state. On the other
hand, for the medium this is a short time; it remains intact, and the
external objects perceive only its averaged (coarse) characteristics.

The external objects are the valence quarks produced by external
currents we pick up at will, and the medium is the QCD vacuum.  The
main problem is the fact that there is no obvious parameter which
could be fine-tuned to ensure a large ratio of two frequencies:  that
of the valence quarks to the typical frequency of the vacuum
fluctuations. We have to rely on numerical parameters whose
existence is not clear {\em a priori}.  Moreover, it may well happen
that in some channels a favorable numerical enhancement exists
while it is
absent in the others.  As a matter of fact, this is exactly what we
have discovered \cite{15}: not all hadrons  (even with small spins)
are alike
the classical $\rho$ mesons or nucleons. There are deep distinctions
in
the hadronic family. This is, probably, the most important
qualitative finding obtained within the sum rule approach, which
escaped
attention of the lattice people. This finding served also as an initial
impetus for the introduction of the instanton liquid
model. I will return to this issue later, and now let me only mention
that the
reasons explaining the
non-universal behavior of different low-lying
hadronic states  are qualitatively understood \cite{15}.

Before dwelling on the technical realization of the idea it is
instructive to discuss the place the method occupies in the catalog of
existing approaches to the hadronic physics. The sum rule method is
admittedly approximate, it requires a certain amount of guesswork
and can
not be formalized in the same sense as, say, the solution of the
Schr{\"o}dinger equation. At the same time, it is not {\em a model}.
Any
model necessarily requires {\em ad hoc} assumptions, and the
accuracy of
the corresponding predictions (estimates) can
not be controlled inside the model itself. In the sum rule method,
once the
rules of the game are accepted (the condensates are introduced once
and
forever) there is no freedom left; they tell you themselves whether
this or
that particular problem is solvable and what accuracy should be
expected.
In a sense, they must be compared to the lattice calculations. The
strength
of the latter is that they can  indefinitely improve with time. The
virtue of the sum rule method is that it is analytic, simple and is
open for the qualitative analysis where one can easily see what
stems from
what.

\section{Implementation of the Idea}

In order to sketch the basic technical elements of the method let us
discuss the problem of charmonium. This example is singled out for
two reasons. First, the theoretical situation here is somewhat simpler
and cleaner than in other cases. The second motivation is
historical (it should be, perhaps, considered as the main at this
Conference, with the  focus on the history of QCD). The first
estimates of the gluon condensate have been obtained from the
charmonium sum rules  [9,16]. The latter work devoted to the $0^-$
charmonium level, $\eta_c$, has a particularly dramatic history.  The
point is that the $\eta_c$ particle was first found experimentally at a
wrong place (2.83 GeV).

The value of the gluon condensate has been
first extracted from the analysis of J/$\psi$, and then applied in the
$0^-$ channel. The result of this investigation was formulated as a
categoric prediction for the $\eta_c$ mass, $(M_{\eta_c})_{theor} =
3.00\pm 0.03$ GeV. The {\em later} discovery of the genuine
$\eta_c$ state
with mass 2.98 GeV \cite{17} was one of the greatest successes of QCD and
the strongest impetus for further work on the sum rules.

The $1^-$ charmonium state, J/$\psi$ particle, is produced from the
vacuum by the current
$$ j_\mu =\bar c\gamma_\mu c $$
where $c$ is the charmed quark field. Now, let us consider the
correlation function
\begin{equation}
\Pi_{\mu\nu}(x) = <0\mid T\{j_\mu (x) j_\nu (0)\}\mid 0> .
\end{equation}
If $x$ is sufficiently small $\Pi_{\mu\nu}$ is given by the sum of the
Feynman graphs like those presented on Fig. 1 where the solid
line is the quark propagator while the dashed line denotes the gluon
propagator.

For heavy quarks the small values of  $x$ do not necessarily imply
that we have to consider  large euclidean momenta in
$\Pi_{\mu\nu}(Q)$.  Indeed, even if $Q\sim 0$ the characteristic
distance between the points of emission of the heavy quarks and
their absorption is of order $x\sim (2m_c)^{-1} \ll\Lambda_{QCD}^{-
1}$.  Moreover, if the loop integrations are saturated in the domain of
large virtual momenta we can use the bare expressions for the quark
and gluon propagators,
$(p^2+m_c^2)^{-1}$ and $(k^2)^{-1}$, respectively.

The fact that the loop integrals are saturated at large virtual
momenta does not mean that there is {\em no} contribution coming
from
the small momenta, $k\leq\Lambda_{QCD}$. It is clear that here,
at $k\leq\Lambda_{QCD}$,
we make a gross mistake by calculating the Feynman graphs with
the
bare propagators.

What can be done under the circumstances? Let us exclude the
domain of small $k$ from the calculation of the perturbative
correction of Fig. 1b.  In order to define what we actually exclude we
will need to introduce the normalization point $\mu$. For $\mid
k\mid >\mu$ we use the bare gluon propagator, for $\mid k\mid
<\mu$ we do not know what to use. Let us observe, however, that
the momentum flowing along the quark line is large, and, hence,
we can approximate $k < \mu$ by $k=0$ in the quark
propagator; what is left of the gluon line is just the integral over the
unknown gluon Green function over the domain $\mid k\mid <\mu$.
Graphically we can depict this contribution as on Fig. 2. In this way
the gluon condensate shows up. The result  obviously reduces to
$$ \frac{<\alpha_s(A_\rho^a A_\rho^a)_\mu >}{m_c^2} f(Q^2/m_c^2)$$
where $f$ is a calculable function determined by the lines which are
far off shell. In the case at hand $f$ vanishes due to the gauge
invariance. The vertex of emission of the soft gluons should contain
$k_\alpha k_\beta$ in the sum of three graphs (Fig. 3). Then the
contribution of Fig. 3 can be written as follows:
\begin{equation}
\frac{<\alpha_sG_{\alpha\beta}^2>_\mu}{m_c^4} F(\frac{Q^2}{m_c^2})
\end{equation}
where $G_{\alpha\beta}$ is the gluon field operator and
the subscript $\mu$ reminds us of the normalization point.  In
the general case any Feynman graph can be prepared in this way,
and
we arrive at the following decomposition
\begin{equation}
i\int e^{iqx}T\{j(x)j(0)\} = \sum_n C_n(Q, m_c,\mu ){\cal O}_n(\mu )
\end{equation}
where the set $\{ {\cal O}_n \}$ includes all local gauge invariant
operators expressible in terms of the gluon fields and the fields of
the light quarks.  Eq. (3) is a concise form of the Wilson operator
expansion. The coefficients $C_n(\mu)$, by construction, include only
the domain $\mid k\mid >\mu$.

So far we have done nothing particularly interesting. The procedure
of separation of the loop momenta described above is quite universal
and is applicable in any theory, say QED, or two-dimensional
$\sigma$ model. It is merely an identical transformation. Moreover,
it looks rather awkward in QED since, instead of directly calculating
the Feynman graphs, it prescribes first to do the separation in the
integrands and then to calculate the very same graphs  in the
separated form.

In QCD this is more than a simple reshuffling. Indeed, we suspect
(actually we are sure) that at $\mid k\mid < \mu$ the Green
functions
are  different from their bare expressions. The details of the infrared
behavior are not known, but a few integrals can be parametrized.
What is absolutely crucial is that the difference between the genuine
propagators and smooth extrapolations from the perturbative
domain is enormous (see Fig. 4). Moreover, the transition from
the perturbative to non-perturbative regime is very sharp. These
two facts are not derivable in the present-day theory; we learned of
them indirectly, by accumulating empiric evidence. They  make the
sum rule approach technically manageable and ensure  predictive
power.

Indeed, under the circumstances the coefficients $C_n$ are {\em
approximately} determined by the standard perturbation theory. The
fact that we have to "cut a hole" in the Feynman graphs (to discard
the domain $\mid k\mid <\mu$ ) is not important numerically since
$\mu^2$ can be chosen sufficiently small in  the
characteristic hadronic scale set by $<4\pi^2G^2>^{1/2}$.
At the same time, under such
choice of $\mu$, the vacuum condensates are due to
non-perturbative effects, to a good approximation.  These effects are
so
violent, that the small perturbative background is totally negligible,
which implies in turn that  the vacuum parameters $<{\cal
O}_n>$ are practically
$\mu$ independent, and there is no need to bother about
the precise definition of the $\mu$-separation procedure (Fig. 4).
Needless to say, that this is  great luck;
in other models the situation may be less favorable. (As an example
where
this element of luck becomes a parametrically exact assertion let me
mention the
$O(N)$ $\sigma$ model. One can show \cite{NSVZ}
that in the leading $1/N$ approximation
the condensate parameters in this model are $\mu$ independent.  At
the same
time, in the next-to-leading  $1/N$ order they  are
very sensitive to the $\mu$ dependence.  A straightforward
generalization
of the QCD sum rules, quite predictive  in the leading $1/N$
approximation, becomes useless at the level of $1/N$.)

Thus  in QCD we arrive  at a pragmatic version of OPE \cite{14}. The
corresponding  simplified prescription reads: calculate the coefficient
functions perturbatively, and hide all non-perturbative effects (and
only
them) in the
vacuum expectation values.  The anomalous dimensions, if present,
should
be taken into account,  of course.  It is worth emphasizing again that
strictly speaking this  prescription is only approximate. The Wilson
OPE does not discriminate perturbative against non-perturbative,
rather it discriminates  small $k$ against large $k$.

Returning to the correlator (1) and applying the approach described
above to the graphs of Fig. 1 (see also Fig. 3) we get
$$\Pi_{\mu\nu}(q) = (q^2g_{\mu\nu} -q_\mu q_\nu )\Pi (q),$$
\begin{equation}
\Pi (q) =F_{0}^{(0)}(q/m_c) +\alpha_sF_{0}^{(1)}(q/m_c)
+\frac{<\alpha_sG^2>}{m_c^4}F_4(q/m_c)
+...
\end{equation}
where $F$ are dimensionless  functions whose subscript indicates the
(normal) dimension of the corresponding operator. Notice that the
$F_{0}^{(1)}$ term contains, generally speaking,  an $O(\mu^4)$ part
whose origin is associated with the constraint in the Feynman
integral.
It must be compensated by a similar  contribution  coming from
$<G^2>$. Both are numerically small and are neglected in eq. (4).

If all terms were known we could find the sum and then the position
of the pole and its residue exactly. The position of the pole is the
$J/\psi$ mass.

Since the vacuum condensates are not calculable (at least, now) we
have to truncate the sum. To keep the theoretical part of  eq. (4)
under control it is desirable to limit ourselves to a few condensate
parameters. The proliferation of these parameters would make the
analysis unmanageable and would be equivalent, in essence,  to a
model dependence. Needless to say, that this was not what we
wanted.

With a few terms in the right-hand side there is no way to reach  the
pole. In other words, the evolution of the $\bar c c$ pair in the
vacuum medium from zero to infinity can not be traced within our
approach.  On the other hand, if the evolution time is not infinite the
correlator $\Pi$ is saturated not only by the lowest-lying state in the
given channel, $J/\psi$, but by excited states as well. Thus, there is a
conflict of interests. To single out the ground state with the given
quantum numbers we need large values of $x$. To reduce the
theoretical calculation to a few vacuum condensates describing the
average vacuum characteristics we need relatively small  values of
$x$. Is there a window where both requirements are met?

The answer to this question is not universal and depends on what
hadronic channel is under consideration. For $J/\psi$, as well as for
classical hadrons like $\rho$ or nucleon, the fiducial domain does
exist, as has been shown in ref. \cite{12}
and later investigations \cite{Ioff}.  Now, by
comparing the theoretical expression for $\Pi (q)$ in the fiducial
domain with its phenomenological counterpart we extract the
parameters of the $J/\psi$ meson.

Since the existence of the window can not be guaranteed by fine-tuning of a
parameter and relies on a subtle numerical balance
different tricks aimed at minimizing the contamination due to the
excited
states and enhancing  the role of the condensates from the standard
set are
very helpful, especially for the light-quark mesons. One such trick is
the borelization procedure suggested in \cite{12}. First, it factorially
suppresses the operators with high dimensions; second, it
simultaneously
transforms the standard dispersion relation in an exponential
integral
where the weight of the excited states is exponentially small. A close
strategy \cite{Shur} in the light-quark channels is  considering
 the correlation
function $\Pi (x) $ directly in the coordinate space, instead of
analyzing
$\Pi (Q)$ in the momentum space. Fig. 5  illustrates the existence of
the
window and the quality of the postdiction for  the $\rho$ meson
mass
and residue. The
vertical axis presents the ratio of the correlation function induced by
the
current $\bar u\gamma_\mu d$
to that for the bare quarks (solid line). The wavy line is the same
ratio
obtained theoretically with the gluon and quark condensates
included. The
arrows $A$ and $B$ show the boundaries of the window (the fiducial
domain).

\section{Current Status}

Unlike QED Quantum Chromodynamics is deprived of a truly small
expansion parameter like $\alpha = 1/137$.  Even $1/N_c$, the last
hope of many theorists, is not that small (I remind that $N_c$ is the
number of colors in the multicolor version of 't Hooft \cite{tHoo}).
Unfortunately (or, perhaps, fortunately) we have to deal with the
genuine strong coupling regime. Moreover, QCD is a notoriously rich
theory  responsible for  an incredible diversity of phenomena in the
hadronic world. Therefore I dare to conjecture that  iterative
procedure allowing one, at least in principle, to calculate the hadronic
parameters with arbitrary accuracy, will not be found in the near
future. It is not high precision which we should  target but, rather,
high reliability of theoretical predictions and full understanding and
control over possible uncertainties.

In this respect the situation with the QCD sum rules, if not ideal, is at
least satisfactory. The spectrum of applications of the method is very
broad now. There are very few problems left in the low-energy
hadronic physics not analyzed within the QCD sum rules. In some
instances the analysis fails to produce reasonable results, and the
challenge is to understand why. What is even more important, the
sum rules can be used in both directions: each new hadronic channel
is a potential source of information about the QCD vacuum; at the
same time whatever new we learn of the vacuum we can
immediately translate this knowledge in new predictions, say, in the
glueball sector.

The standard set of the condensates which are in use now
essentially reduces to  five: two gluon condensates, two quark and
one mixed,
$$ <\alpha_sG^2>, \,\, <g^3G^3>,$$
$$<\bar\psi \psi >, \,\, <\bar\psi \Gamma\psi \bar\psi \Gamma\psi
>,$$
\begin{equation}
<\bar\psi \sigma_{\mu\nu}G_{\mu\nu}\psi > .
\end{equation}
True, these parameters enter the sum rules on uneven footing.  The
cubic gluon condensate, as a rule, plays an insignificant role. The
quartic quark condensate is not treated as independent. Instead,
invoking factorization (which is, in turn, justified by large $N_c$
arguments) one reduces it to the square of $<\bar\psi \psi >$.
Although it is quite obvious that
deviations from factorization should be present at a certain level
numerous attempts to detect them theoretically
yielded no conclusive evidence. Finally the
mixed quark-gluon condensate shows up at a noticeable  level
mainly
in the baryon sum rules. As a matter of fact, its numerical value has
been extracted first from the analysis in the nucleon channel
\cite{Bely}.

Using essentially three phenomenological numbers plus the quark masses
as input the QCD
practitioners were able to understand an enormous wealth of data
referring to the low-energy hadronic physics.
For this Conference I intended to prepare  a list of
the most spectacular predictions  (postdictions) illustrating this
statement. It turned out impossible
for two reasons. First, it became rather boring. Whatever parameter
you randomly choose from the Rosenfeld Tables
(say, the $\rho$ meson mass, or $g_A$ --
the nucleon axial constant -- or the pion charge radius, or...) you look in
the
Tables, then in the corresponding papers, and then you find total
agreement within the expected theoretical accuracy.
More seriously, the number of works reporting on
remarkable achievements is so large that making a short list of "most
important" results would, probably, be unfair, the more so that
all results quoted would require comments, and the "short list"
would become not so short very soon.
A general idea of the range of problems solved can be obtained from
the review paper \cite{RRY} and the
Reprint
Volume \cite{MS}. It would be quite in order to have a book covering
different versions of the method existing today and  all major
applications. Alas, such book is not written yet. The best I can do now is
to try to catalog the main directions in which the sum rule method has been
developing.

It started from the masses and residues of the low-lying hadronic states.
All conceivable combinations -- light quarks, quarks and gluons (exotics),
glueballs, heavy quarks, light and heavy quarks; mesons and baryons -- are
considered and analyzed. At the second stage it was understood how to treat
more complicated static properties, such as magnetic moments, charge radii,
the axial and vector coupling constants for baryons, etc. Form factors at
intermediate values of the momentum transfer and the light-cone wave
functions were next logical steps. In the mature phase the QCD sum rules
were applied to very sophisticated processes which are so far beyond reach
of other approaches. The nucleon structure functions and the weak non-leptonic
and radiative decays are just two examples immediately coming to
one's mind.  I bring my apologies to my colleagues for the fact
that I failed to compile even an
incomplete list of references at this point.  The
interested reader might turn to the Reprint Volume \cite{MS}.

Surprisingly, the method is applied now  far beyond the scope it was
originally designed for. Let me mention, for instance, recent attempts
to expand it to cover the cases of the nuclear matter \cite{nucl}
  and/or non-zero
temperatures and densities \cite{Shur}.  This is a promising although
complicated direction where the theoretical situation is so volatile
that I do not risk  to discuss it here in more detail.  A less
unexpected field where the sum rules successfully compete
\cite{Neub} with
other approaches is the Effective Heavy Quark Theory.

As it often happens  the most intriguing problems -- the epicenter of
the current efforts -- are those where the sum rule approach fails.
One failure is quite obvious: the large-spin hadrons \cite{Shi}.
Indeed, the
latter have parametrically large sizes and a "sausage-like shape"
(growing with spin) and, therefore, it is quite clear that the basic idea
of the method -- extrapolation from short to intermediate distances
--  is not applicable. Practically we have to stop at $S=2$. Higher
spins
are the realm of stringy models or theories (if and when they are
created, of course).

Even for small spins  there exist clear-cut examples  where the
interaction of the valence quarks and gluons with the vacuum
fluctuations is so peculiar  (and strong)  that the average vacuum
characteristics are not enough to adequately describe how the
corresponding hadrons are formed.  Such a situation occurs in
$0^{\pm}$ quark mesons and, especially in  $0^{\pm}$ glueballs
\cite{15}.
In these channels it is absolutely essential to know the details of the
vacuum fluctuations, not just average densities.  Hence, they will
serve as the most sensitive testing ground for any future
"fundamental" theory of hadrons. It is remarkable -- and I would
like to emphasize it again -- that in these
channels we find a new scale in the physics of low-lying states,
and this scale is significantly larger than that we got used to in the
classical mesons and baryons. It is also worth noting that these are
just the channels where the $1/N_c$ counting also fails.
\footnote{Let me parenthetically mention a problem where
the sum rules are supposed to
work very well, and still the corresponding prediction lies a factor of 3
higher than the experimental number, far beyond the error bars. I mean the
$J/\psi \rightarrow \eta_c\gamma$ decay. I would bet that the theory is right
and experiment wrong.}

The best
what can be done to-day is to develop particular models of the QCD
vacuum. At least two of them are on the market: the instanton liquid
model \cite{Shur2} and the stochastic vacuum model \cite{Dosch}.
Both have been
originally motivated by the QCD sum rules
but are much more advanced in specifying  the  dominant vacuum
fields.  According to \cite{Shur2} the typical vacuum fields are given,
in a
reasonable approximation, by distorted instantons which still do not
loose their individuality. The second model postulates the dominance
of the stochastic type fluctuations with a finite correlation length.
The detailization of the vacuum structure is  the strength of these
models -- they are applicable even in those problems where there is
no hope of using the standard QCD sum rules --  but, simultaneously,
this is their weakness.  The concrete assumptions  the authors had to
make are not derivable so far from first principles rendering the
corresponding results model-dependent and vulnerable to all sorts of
critical remarks scattered in the literature.

There  is something mysterious in the instanton-liquid model.
On one hand, if you will attentively follow the model and do the
calculations you will observe that literally speaking it leaves no
space
for the window
like on Fig. 5. The quark and gluon condensates come out so large
that they
enter  the game before the contamination dies off. On the other hand,
the
full curve  which you could get in the model nicely repeats our
theoretical
prediction up to $\sim$ 1.5 fm!  It is clear that this aspect calls for an
immediate explanation.

I have to add that there is an ongoing controversy in the current
literature over several aspects of the QCD sum rule approach. Such
issues as the value of the basic gluon condensate, factorization, the
status of the pragmatic version of OPE, the possible role of higher
condensates, etc. are vividly debated. This controversy is not
necessarily a negative fact. On the contrary, it shows that the method
is still in its active stage, the development goes on, new ideas
continue to appear -- naturally -- along with new question marks.
Unfortunately, I have no time here to dwell on the essence of the
arguments and explain, say, why I believe that the original estimate
of the gluon condensate from  charmonium is quite accurate or why
factorization should hold.  Those who are interested can look through
my comments in the Reprint Volume \cite{MS}.

\section{A Recent Example}

In this part of my talk, just
to amuse the audience and illustrate the virtues
and drawbacks of the method, I will show how it works in  an
almost back-of-the-envelope
estimate of the non-factorizable terms in the weak non-leptonic
decays
\cite{Blok}. The estimate is not very precise; hopefully, it is valid up
to
a factor of two, but it is a very important phenomenological
issue, and I am sure it will take quite a while before something
like that appears on the lattice.

Let me remind you that the non-factorizable terms in the decays like
$\bar B^0 \rightarrow D^+\pi^-$ are reducible to the amplitude
\begin{equation}
<D\pi\mid {\tilde{\cal O}}_2\mid \bar B >
\end{equation}
where
\begin{equation}
{\tilde{\cal O}}_2 = 2(\bar c\Gamma_\mu t^a b)(\bar d \Gamma^\mu
t^a u),
\end{equation}
the tilde appears due to historical reasons, $t^a$ is the color
matrix
while $\Gamma$ is the $V-A$ matrix. Within the naive factorization
this
amplitude vanishes. Of course, the perturbative gluon exchange
between the
heavy and the light quark brackets eliminates this zero, but the
result is so
small numerically that it can be safely neglected.

Still,  color has to be exchanged between the brackets because
otherwise
the light quarks can not form the pion. The job has to be done by a
soft
gluon.

To estimate the effect let us consider the correlation function
\begin{equation}
{\cal A}^\beta = \int d^4x <D\mid T\{ {\tilde{\cal O}}_2(x), A^\beta (0)
\} \mid \bar B > e^{iqx}
\end{equation}
where $A^\beta$ is an auxiliary axial current annihilating the pion,
$$ A^\beta = \bar u\gamma^\beta \gamma^5 d $$
and $q$ is an external momentum flowing through $A^\beta$. We
assume it to
be euclidean, neither too small nor too large. (This is the
intermediate
domain we always speak about). Now, let us  consider for simplicity
the limit in which the masses of the D and B mesons are close to each
other,
an assumption  which is very helpful technically.
Then the
correlator (8) is trivially calculable,
\begin{equation}
{\cal A}^\beta = -\frac{i}{4\pi^2}\frac{q^\alpha q^\beta}{q^2}
<D\mid \bar c\Gamma^\mu t^a g {\tilde G}^a_{\alpha\mu} b\mid B>
\end{equation}
plus terms suppressed by powers of $1/q^2$. Here ${\tilde
G}^a_{\alpha\mu}$
is the (dual) gluon field strength tensor.

For the  express evaluation we omit the terms of the higher order in
$q^{-
2}$ on the right-hand side. The only operator retained describes the
effect
of the soft gluon absorption somewhere in the gluon medium
surrounding the
heavy quark in the corresponding heavy meson. If the heavy quarks
are
treated non-relativistically then one component of the operator
dominates
over the others, namely, $\alpha =0$ (in the quark rest frame) and
$\Gamma^\mu
\rightarrow \gamma^i\gamma^5$. In this approximation
\begin{equation}
\bar c\Gamma^\mu t^a g {\tilde G}^a_{\alpha\mu} b \rightarrow
-g\vec\sigma {\vec H}^a t^a,
\end{equation}
where $\vec\sigma$  represents the Pauli spin  of the heavy quark
while the
${\vec H}^a $ is the chromomagnetic field operator.

Although we are unable at the moment to calculate the average
chromomagnetic field inside heavy mesons from first principles, it is
--
nevertheless -- known phenomenologically, an analog of the gluon
condensate
(5).  Indeed, $g\vec\sigma {\vec H}^a t^a$
is the leading term splitting the masses of $B^*$ and $B$. Hence, the
matrix element on the right-hand side of eq. (9) reduces to
$$m_{\sigma H}^2 \equiv \frac{3}{4}(M^2_{B^*}-M^2_B).$$

Saturating eq. (9) by the pion contribution we find
\begin{equation}
<D^+\pi^-\mid {\tilde{\cal O}}_2\mid {\bar B}^0 >\sim
\frac{i}{4\pi^2f_\pi}m_{\sigma H}^2(M^2_B-M^2_D).
\end{equation}
It is convenient to normalize this prediction to the factorizable part
associated with ${\cal O}_2$. The ratio of the non-factorizable to
factorizable parts is \cite{Blok}
\begin{equation}
r \sim -\frac{N_cm_{\sigma H}^2}{4\pi^2f_\pi^2} \sim -1.
\end{equation}
It is worth noting that the expression for $r$ quoted above is
$O(N_c^0)$
since $f_\pi^2 \sim N_c$, as was expected, of course.
The result (12) will set the scale for all future estimates of the
deviations
from factorization which, eventually, will have better accuracy.

\section{Conclusions}

Sometimes people say that only in the hard processes -- like the
logarithmic
evolution of the structure functions -- QCD is a true success while in
the
realm of the genuinely hadronic physics it has not shed much light.
Very
few facts which were totally beyond our comprehension before 1972
are
understood now analytically.

This statement is correct only in part. Yes, the {\em full}
analytic solution of the soft hadronic physics is still lacking.
However, the QCD sum rules do provide a new insight, both
in quantitative and qualitative issues. They are extremely successful
in
correlating essentially every parameter from the Rosenfeld tables,
dozens
of them, to three or four vacuum expectation values. They present a
useful,
and in many instances, unique tool for reliable estimates in the low-
energy
domain. They predict the occurrence of a new large scale in certain
channels
with the vacuum (or "almost" vacuum) quantum numbers.
Certainly, they do not explain how the
infrared soup is cooked,  but taking this fact for granted, they
skillfully utilize the  recipe.

The QCD sum rules are engineered as an approximate computational
scheme in
the regime of strong coupling -- QCD at intermediate distances. The
method
can not be used iteratively, in order to achieve arbitrary accuracy,
like
the $\alpha$ expansion in QED. What is important, however, is the
fact that the
accuracy of the approximations done can be controlled within the
method
itself. Using this approach we get a qualitative idea of the structure
of
the QCD vacuum which will stay with us irrespectively of  further
developments. This idea is immediately translatable in predictions
concerning new regularities in the hadronic family.

Finally, let me notice that the method is very flexible and can easily
accommodate new information appearing on the market, as was, for
instance,
the case with the large $N_c$ expansion. It can be used in
conjunction with
other approaches, both analytic and numerical (heavy quark
expansion,
lattice calculations, etc.).

\vspace{0.5cm}

{\bf Acknowledgements}

\vspace{0.3cm}

This work was supported in part by DOE under grant number DOE-AC02-
83ER40105.

\end{document}